\documentclass{elsart}

\usepackage{graphicx}

\usepackage{amssymb}

\usepackage{lineno}

\begin{document}
\newcommand {\ve} [1] {\mbox{\boldmath $#1$}}
\newcommand  {\beq} {\begin{eqnarray}}
\newcommand {\eeq} {\end{eqnarray}}
\newcommand{\dem} {\mbox{$\frac{1}{2}$}}
\newcommand {\arrow} [2] {\mbox{$\mathop{\rightarrow}\limits_{#1 \rightarrow #2}$}}

\begin{frontmatter}



\title{Gamma-delayed deuteron emission of the $^6$Li$(0^+;T=1)$ halo state}
\author[pntpm,tas]{E.M. Tursunov},
\ead{tursune@inp.uz}
\author[pntpm]{P. Descouvemont\thanksref{fnrs}}, and 
\ead{pdesc@ulb.ac.be}
\author[pq,pntpm]{D. Baye}  
\ead{dbaye@ulb.ac.be}
\address[pntpm]{Physique Nucl\'eaire Th\'eorique et Physique Math\'ematique, C.P. 229,
Universit\'e Libre de Bruxelles, B1050 Brussels, Belgium}
\address[tas]{Institute of Nuclear Physics, Uzbekistan Academy of Sciences, 702132, Ulugbek, Tashkent, Uzbekistan}
\address[pq]{Physique Quantique, C.P. 165/82,  
Universit\'e Libre de Bruxelles, B1050 Brussels, Belgium}
\thanks[fnrs]{Directeur de Recherches FNRS}

\begin{abstract}
M1 transitions from the $^6$Li($0^+;T=1$) state at 3.563 MeV to the $^6$Li($1^+$) ground state and
to the $\alpha+d$ continuum are studied in a three-body model. 
The bound states are described as an $\alpha+n+p$ system in hyperspherical coordinates 
on a Lagrange mesh.  The ground-state magnetic moment   
and the gamma width of the $^6$Li(0$^+$) resonance are well reproduced.  The halo-like structure 
of the $^6$Li$(0^+)$ resonance is confirmed and is probed by the M1 transition probability to the $\alpha+d$ continuum. The spectrum is sensitive to the description of the $\alpha+d$ phase shifts.
The corresponding gamma width is around 1.0 meV, with optimal potentials. Charge symmetry is analyzed through a comparison with the $\beta$-delayed deuteron spectrum of $^6$He.
In $^6$He, a nearly perfect cancellation effect between short-range and halo contributions was found.
A similar analysis for the $^6$Li($0^+;T=1$) $\gamma$ decay is performed; it shows that charge-symmetry breaking at large distances, due to the different binding energies
and to different charges, reduces this effect.
The present branching ratio
$\Gamma_{\gamma}(0^+\rightarrow \alpha+d)/\Gamma_{\gamma}(0^+\rightarrow1^+)\approx 1.3\times 10^{-4}$ 
should be observable with current experimental facilities.
\end{abstract}

\end{frontmatter}

\section{Introduction}
Electromagnetic transition processes provide a useful tool for the study of the nuclear structure 
and of the reaction mechanisms. The theoretical study of such processes yields estimates for 
the different static and dynamical observables of a nucleus.  
In $^6$Li, the $(0^+; T=1)$ state has raised interest as a good candidate 
for observing parity violation \cite{cso96,gri98}. 
Indeed, its decay into the $\alpha + d$ continuum is forbidden by parity 
conservation. 
Since electromagnetic M1 transitions into this continuum are allowed, 
they have been also studied because they may compete with the parity-violating 
decay and make its detection difficult. 

However, the $^6$Li$(0^+; T=1)$ state is also interesting by itself. 
It is most likely a halo state, as it is the isobaric analog 
of the $^6$He ground state \cite{SY91}. 
M1 transitions to the continuum are in fact also an excellent tool to explore 
these halo properties and compare them with those of $^6$He. 
Recent experimental and theoretical works on the delayed
$^6$He $\beta$ decay suggest that the deuteron spectrum
is strongly sensitive to the halo structure (see Ref.~\cite{TBD06} and references therein).
Similarities between this process and the $\gamma$-delayed
deuteron emission of $^6$Li$(0^+)$ are expected, and should test charge symmetry in exotic light nuclei.

The branching ratio of the total transition probability to the $\alpha+d$ continuum and the
transition probability to the $^6$Li$(1^+)$ ground state was estimated  as $8\times 10^{-5}$ under a number of 
simplifying assumptions \cite{gri98}. However the shape and magnitude of the transition 
probability to the continuum as a function of the deuteron energy were not studied. In addition,
the sensitivity with respect to the $\alpha+d$ potential, as well as convergence problems, were
not addressed.
The aim of the present work is to investigate M1 transitions from the
$^6$Li$(0^+)$ excited state to the $\alpha+d$ continuum, as well as to the $^6$Li$(1^+)$ ground state.
For the 
description of the $^6$Li states, we use different two-body and three-body models. Three-body hyperspherical 
wave functions \cite{ZDF93} are based on the Lagrange-mesh method and give an accurate solution of the three-body Schr{\"o}dinger equation \cite{DDB03}.    
The $\alpha+d$ scattering wave function is factorized into a deuteron 
wave function and a nucleus-nucleus scattering state. This work extends our previous study on $^6$He $\beta$ decay where the same formalism was used.

The model is presented in Section 2. The potentials and the corresponding two-body and  three-body wave 
functions are also described. In Section 3, we discuss the results in comparison with the experimental data, and analyze the sensitivity with respect to the
$\alpha+d$ potential. 
Finally, conclusions are given in Section 4.        
     
\section{Model}
\subsection{Three-body wave functions of $^6$Li bound states}
The $^6$Li bound-state wave functions are defined in an $\alpha+n+p$ model using the hyperspherical 
coordinates \cite{ZDF93}. 
A set of Jacobi coordinates for three particles with mass numbers $A_1=1$, $A_2=1$, and $A_3=4$ 
is defined as 
\beq
\ve{x} =  \sqrt{\mu_{pn}} \, \ve{r}, \hspace*{1 cm}
\ve{y} =  \sqrt{\mu_{\alpha d}} \, \ve{R},
\label{eq206}
\eeq
where the (dimensionless) reduced masses are given by $\mu_{pn}=1/2$ and $\mu_{\alpha d}=4/3$. 
The $p-n$ relative coordinate and the coordinate between $\alpha$ and $d$ are denoted by
$\ve{r}$ and $\ve{R}$, respectively.
Equations (\ref{eq206}) define six coordinates which are transformed to the hyperspherical coordinates as
\beq
\rho^2 = x^2+y^2, \hspace*{1 cm} 
\alpha = \arctan (y/x),
\label{eq207}
\eeq
where $\alpha$ varies between 0 and $\pi/2$. 
With the angular variables $\Omega_x = (\theta_x,\varphi_x)$ and $\Omega_y =(\theta_y,\varphi_y)$, 
equations (\ref{eq207}) define a set of hyperspherical coordinates 
which are known to be well adapted to the three-body Schr{\"o}dinger equation. 

We define $\gamma=(\ell_x,\ell_y,L,S)$ where $\ell_x$ and $\ell_y$ are the orbital momenta
associated with the Jacobi coordinates $\ve{x}$ and $\ve{y}$, respectively. 
With the notation $\Omega_5 = (\alpha, \Omega_x, \Omega_y)$, a three-body wave function
with spin $J$ and parity $\pi$ reads \cite{DDB03} 
\beq
\Psi^{JM\pi}_{^6{\rm Li}}(\rho,\Omega_5) =  \rho^{-5/2}
\sum_{\gamma K} {\chi}^{J\pi}_{\gamma  K}(\rho) 
{\cal Y}^{JM}_{\gamma  K}(\Omega_5),
\label{eq210}
\eeq
where ${\cal Y}^{JM}_{\gamma  K}(\Omega_5)$ are the hyperspherical functions (including spin), defined as
\beq
&&{\cal Y}^{JM}_{\gamma  K}(\Omega_{5})=\phi^{\ell_x \ell_y}_{K}(\alpha) 
\left[ \left[ Y_{\ell_x}(\Omega_{x})\otimes Y_{\ell_y}(\Omega_{y}) \right]^L \otimes {\chi}^S \right]^{JM},   
\nonumber \\
&& \phi^{\ell_x \ell_y}_{K}(\alpha)=
{\cal N}_K^{\ell_x \ell_y} (\cos \alpha)^{\ell_x} (\sin \alpha)^{\ell_y}
P_n^{\ell_y+\dem,\ell_x+\dem}(\cos 2\alpha),
\label{eq210b}
\eeq
with $n=(K-\ell_x-\ell_y)/2$, and where ${\cal N}_K^{\ell_x \ell_y}$ is a normalization factor,
$P_n^{a,b}$ a Jacobi polynomial and ${\chi}^S$ a spin function. 
Details are given in Refs.~\cite{ZDF93,DDB03}. The hyperradial functions ${\chi}^{J\pi}_{\gamma  K}$ are  
obtained from a set of coupled equations truncated at $K=K_{max}$. The problem is solved by using the Lagrange-mesh
technique (see Ref.~\cite{DDB03} for details). 

The three-body wave functions contain components with total intrinsic spin $S=0$ and $S=1$. Because of the positive parity, $\ell_x+\ell_y$ is even and only even $K$ values are involved.

\subsection{$\alpha+d$ two-body wave functions}
As it was done in Ref.~\cite{TBD06} for the $^6$He $\beta$ decay, the scattering $\alpha+d$ 
wave functions are factorized into a deuteron ground-state
wave function, calculated with an appropriate NN potential, and an $\alpha +d$ wave function 
derived from a potential model. We neglect the small $D$ component of the deuteron. 
In the $\alpha+d$ exit channel, only $S$ waves are involved. 
Consequently, the final $1^+$  wave function reads
\begin{equation}
\Psi^{1M+}_{\alpha d}(E,\ve{r},\ve{R})=\Psi_d(\ve{r})\, \Psi_{\alpha d}(E;\ve{R}).
\label{eq200}
\end{equation} 
The spatial part of the deuteron wave function is written as
\beq
\Psi_d(\ve{r})=r^{-1} \, u_d(r)\, Y_{00}(\hat{\ve{r}}) .  
\label{eq200a}
\eeq

The $S$-wave component of the $\alpha+d$ relative motion wave function is factorized as  
\begin{equation}
\Psi_{\alpha d}(E;\ve{R})= R^{-1} \, u_E(R) \, Y_{00}(\hat{\ve{R}}).
\label{eq201}
\end{equation}
The normalization of the scattering wave function is fixed by the asymptotic behaviour as 
\begin{equation}
u_E(R) \arrow{R}{\infty}  F_0(k_{\alpha d}R) \cos\delta_0(E) + G_0(k_{\alpha d}R) \sin\delta_0(E),
\label{eq202}
\end{equation} 
where $F_0$ and $G_0$ are the Coulomb functions, $\delta_0(E)$ is the $s$-wave phase shift at energy $E$,
and $k_{\alpha d}$ is the wave number of the relative motion. 

The present two-body model can also be applied to $\alpha+d$ bound states. In that
case, the scattering wave function $u_E(R)$ in Eq. (\ref{eq201}) is replaced by an $S$-wave bound-state radial function.
 
\subsection{Transition probability per time and energy units}
For the M1 transition to the ground state, the gamma width 
is calculated from
\begin{equation}
\Gamma_{\gamma}(0^+ \to 1^+)= \frac{16 \pi}{9} \, k_{\gamma}^3 \,
 |\langle \Psi^{1^+}_{^6{\rm Li}} 
|| {\cal M}^M_1 || \Psi^{0^+}_{^6{\rm Li}}  \rangle |^2 ,
\label{eq205}
\end{equation}
where  $k_{\gamma}$ is the wave number of the emitted photon.
This definition involves bound-state wave functions on both sides.

With the normalization (\ref{eq202}) of the scattering wave function, the M1 transition probability 
of the process
\begin{equation}
^6{\rm Li(0^+)} \, \to  \, \alpha \, + \, d\,+ \gamma,
\label{eq203}
\end{equation}     
per time and energy units, is given by reduced matrix elements between the initial bound state and the final
scattering states as  (see Appendix A)
\begin{equation}
\frac{dW_{\gamma}}{dE}= \frac{32 \mu_{ad} m_N }{3 \hbar^3 k_{\alpha d}} \, k_{\gamma}^3\,
|\langle \Psi^{1^+}_{\alpha d}(E)|| {\cal M}^M_1 ||\Psi^{0^+}_{^6{\rm Li}}  \rangle |^2 ,
\label{eq204}
\end{equation}
where $m_N$ is the nucleon mass.
The maximum $\alpha+d$ energy is $Q=2.089$ MeV.
The M1 differential gamma width per energy unit to continuum states is expressed as
\begin{equation}
\frac{d\Gamma_{\gamma} (0^+ \to \alpha+d)}{dE}= \hbar \,\frac{dW_{\gamma}}{dE} ,
\label{eq2044}
\end{equation}
and the total width is deduced by integration over the energy.

The M1 operator contains orbital and spin-dependent components.
For a general three-body system, it reads, in Jacobi coordinates \cite{DDB03}
\beq
{\cal M}^M_{1 \mu}(\ve{x},\ve{y}) &= \mu_N \sqrt{\frac{3}{4\pi}} 
[& A_x \ell_{x,\mu}  + A_y \ell_{y,\mu} + 
A_{xy} (\ve{x}\times \ve{p_y}+ \ve{y}\times \ve{p_x})_{\mu}  \nonumber \\
 && + \sum_{i=1}^3 g_s(i)  {\ve{s}}_{i\mu}],
\label{M1oper}
\eeq
where $\mu_N=e\hbar/m_N c$ is the nuclear magneton, ${\ve{s}}_{i}$ are the spins of the three particles, and
$g_s(i)$ their gyromagnetic factors. Coefficients $A_x$, $A_y$ and $A_{xy}$ are related to the mass 
and charge numbers as
\beq
&&A_x=\frac{Z_2 A_1^2+Z_1A_2^2}{A_1A_2 A_{12}}, \nonumber \\
&&A_y=\frac{(Z_{1}+Z_2)A_3^2+Z_3A_{12}^2}{AA_{12}A_3}, \nonumber \\
&&A_{xy}=\sqrt{\frac{A_1A_2A_3}{A_{12}^2A}}\left( \frac{Z_1}{A_1}-\frac{Z_2}{A_2} \right ),
\label{coef}
\eeq
where $A_{12}$ is the reduced mass of the $1+2$ system; in the present case it is denoted as $\mu_{pn}$.
Variables $\ve{p_x}$ and $\ve{p_y}$ are the momenta associated 
with the Jacobi coordinates $\ve{x}$ and $\ve{y}$, respectively.
The matrix elements of the M1 operator between hyperspherical functions are given in Appendix B.

\section{Results and discussion}
\subsection{Conditions of the calculations}
\subsubsection{Three-body wave functions of the $1^+$ and $0^+$ states}
The initial $0^+$ wave function is calculated in an $\alpha+n+p$ three-cluster model,
using hyperspherical coordinates, 
as explained in Ref.~\cite{DDB03}. The same model is applied to the $^6$Li ground state.
In both cases, the Coulomb $\alpha+p$ interaction is included, and is taken as a point-sphere potential parameterized as $V_C(r)=2e^2 {\rm erf}(r/R_C)$ with a radius $R_C=1.2$ fm.
Two-body forbidden states are removed by using the Orthogonalising Pseudopotential method \cite{KP78}.
The central Minnesota interaction \cite{thom77} describes the $n+p$ system. It is adjusted on
the deuteron binding energy 
and reproduces fairly well nucleon-nucleon phase shifts at low energies. 
For the $\alpha+N$ nuclear interaction we employ the potential of Voronchev {\sl et al.} \cite{vor95},
slightly renormalized by a scaling factor (1.008 for $1^+$ and 1.043 for $0^+$) to
reproduce the experimental energies with respect to the three-body threshold ($-3.70$ MeV for the ground state,
and $-0.13$ MeV for the $0^+$ state). 
We truncate the hypermomentum expansion to $K_{max}=20$ which 
ensures a good convergence of the energies.

\begin{figure}[thb]
\begin{center}
\includegraphics[width=10cm]{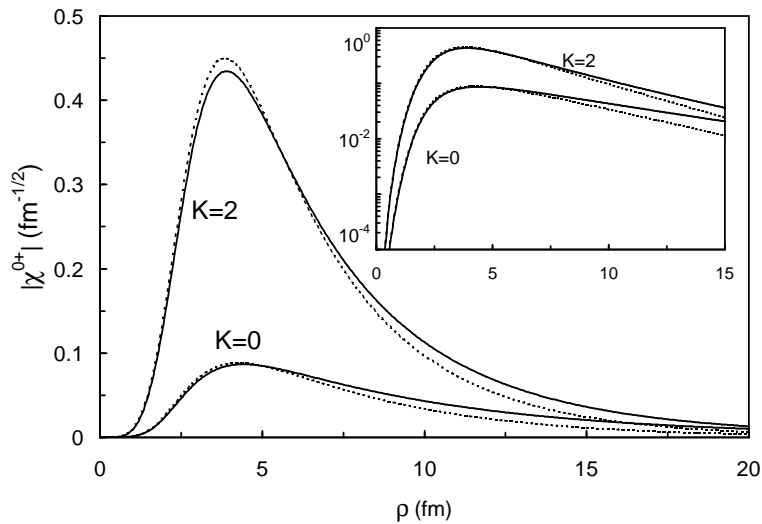}
\end{center}
\caption{Hyperradial wave functions for $^6$Li($0^+)$ (solid lines) and $^6$He (dashed lines) corresponding
to $S=L=0,K=0,2$. The insert shows the same plot in a logarithmic scale. 
\label{Fig0}}
\end{figure}

The matter r.m.s. radius of the ground state (with 1.4 fm as $\alpha$ radius) is found as 
$\sqrt{<r^2>}=2.25$ fm, a value slightly lower than the experimental value ($2.32\pm 0.03$ fm \cite{TKY88})
(note however that a significantly larger radius, $2.54\pm 0.03$ fm, was found in Ref.~\cite{PMP06}). 
For the excited
$0^+$ level, we find $\sqrt{<r^2>}=2.56$ fm, which is close to 
the $^6$He radius. This large value confirms the halo structure of this state \cite{ASV95}. The ground state is essentially $S=1$ (96.0\%). The  $S=0$ component is 84.4\% for $^6$Li($0^+$) and 82.1\% for $^6$He. 
The $^6$Li($0^+$) and $^6$He hyperradial wave functions are plotted in Fig.~\ref{Fig0} for the dominant $K=0,2$
hypermoments. The overlap $\langle^6 \textrm{He}|T^+/\sqrt{2}|^6\textrm{Li}(0^+)\rangle=0.996$ is in good agreement with the value of 
Arai {\sl et al} \cite{ASV95} (0.995).
According to charge symmetry the short-range parts of the $^6$He
and $^6$Li($0^+$) analog levels should be very close to each other.  This is confirmed by Fig.~\ref{Fig0}.
On the contrary, the halo components of
both wave functions are expected to differ significantly: the charges of the halo nucleons are 
different, and the binding energy of $^6$Li($0^+$) is much lower. Consequently, the asymptotic
decrease of the wave function is slower, and matrix elements involving this long-range part should
be different from their analogs in $^6$He.

\subsubsection{$\alpha+d$ scattering states}
In the following we use four different $\alpha+d$ potentials (the phase shifts can be found in 
Ref.~\cite{TBD06}):
\begin{enumerate}
\item The attractive Gaussian potential of Ref.~\cite{Dub94} $V_a$ contains a forbidden state 
and provides the correct $^6$Li binding energy  
($E_1=-1.473$ MeV, with respect to the $\alpha+d$ threshold). Owing to the presence of a forbidden state, it also provides a good fit of the low-energy experimental phase shifts.
\item To test the influence of the short-range part of the $\alpha+d$ wave functions, and 
in particular of the node location, we also use the potential $V_{a}^{S1}$, obtained 
from a supersymmetric transformation \cite{Ba87}. The resulting potential gives the same phase
shifts and the same ground-state energy as the initial potential, 
but the forbidden state is removed and the role of the Pauli principle is simulated by a short-range core.
\item The calculation is complemented by two folding potentials, using  the deuteron wave function $\Psi_d$  provided by the Minnesota potential. These potentials present one forbidden state.
The folding potential $V_{f1}$ is obtained from the original
$\alpha+N$ potential with a renormalization factor 1.068, which yields the correct binding energy for 
$^6$Li; however the quality of the $S$-wave phase shift is poor.   
The folding potential $V_{f2}$ with a renormalization factor 1.15 describes the $S$-wave phase shift 
accurately, but overestimates the binding energy of the $^6$Li ground state ($-2.386$ MeV).
\end{enumerate}

In all cases, the $\alpha+d$ Coulomb potential is chosen as in Ref.~\cite{Dub94}, i.e. as a
bare Coulomb potential.

\subsection{$M1$ properties of bound states}
A test of three-body wave functions is provided by M1 spectroscopic properties, which are well known 
experimentally \cite{TCG02}. In Table~\ref{table1}, we present the calculated values of the 
magnetic moment and of the B(M1) in $^6$Li. Separate contributions are given for the orbital
and spin terms 
of the M1 operator [see Eq.~(\ref{M1oper})]. In both cases, the contribution of the orbital
term is small since the dominant component in the ground-state wave function is an $S$ wave.
The main contribution to the M1 matrix element comes from the spin term. 
The present matrix element corresponds to $B(M1)=7.9$ W.u., or $\Gamma_{\gamma}=7.5$ eV, which are in
good agreement with experiment ($8.62 \pm 0.18$ W.u. and $8.19\pm 0.17$ eV, respectively).
The results are also close to those of Kukulin {\sl et al.} \cite{KPR95} who use different variants of a three-body model.

\vspace*{1cm}
\begin{table}[h]
\caption{Contributions (in $\mu_N$) of the orbital $(L)$ and spin $(S)$
components to the M1 matrix elements. The three-body model is used for the $0^+$ state. 
Experimental data are taken from \cite{TCG02}. }
\begin{tabular}{lcccc} 
\hline
            &  $(L)$ & $(S)$  & Sum  & Exp. \\ \hline
\underline{Three-body model for $1^+$} \\ 
$\mu(^6$Li) &   0.02 &0.84 & 0.86 & 0.82  \\  
$\langle \Psi^{1^+}_{^6{\rm Li}} 
|| {\cal M}^M_1 || \Psi^{0^+}_{^6{\rm Li}}  \rangle$ 
  &  0.13 &2.04 & 2.17 & 2.28 \\ 
\underline{Two-body model for $1^+$} \\           
$\mu(^6$Li) &          0                  & 0.88  & 0.88 & 0.82  \\  
$\langle \Psi^{1^+}_{^6{\rm Li}} || {\cal M}^M_1 || \Psi^{0^+}_{^6{\rm Li}}  \rangle$ 
    & 0.04 &1.53  &1.57 & 2.28 \\  
 \hline
\end{tabular}
\label{table1}
\end{table}
\vspace*{1cm}

These matrix elements can also be obtained with a 2-body description of the $^6$Li ground state. In that case
we use the potential $V_a$ to generate the wave functions.
Since components with $L\neq 0$ are small in the ground-state wave function, the two-body model
is expected to be a good approximation. The $1^+$ magnetic moment in the two-body model is a simple sum of the proton and neutron 
magnetic moments. In this case, both approaches provide similar results, in good agreement with experiment.
However the rms radius in the two-body model is $\sqrt{<r^2>}=2.11$ fm, lower than experiment and than
the three-body value (see Sect.3.1.1).
In addition, the M1 transition matrix element (see Table \ref{table1})
provides $\Gamma_{\gamma}=3.9$ eV, i.e. an underestimate of the experimental value. These results suggest
that the short-range part of the two-body description is too simple. However, transitions to the
continuum are more sensitive to the long-range part of the $\alpha-d$ wave functions.

\subsection{M1-transition to the $\alpha+d$ continuum: effective wave functions and their integrals}
Since the $\alpha+d$ relative motion is described by $S$ waves, 
the first and second orbital terms of the M1 transition operator (\ref{M1oper}) 
do not contribute to the reduced
matrix elements for transitions to the $\alpha+d$ continuum. The orbital 
and spin terms yield nonzero matrix elements only for the 
$\ell_x=\ell_y=L=S=1$ and $\ell_x=\ell_y=L=S=0$ 
components of the three-body wave function, respectively. 
As it will be shown further, the main contribution comes from the spin part of the 
transition operator. The $P$-wave hyperspherical components give small corrections to the process
since the $0^+$ state is essentially $S=0$. 

In order to analyze the $\gamma$-decay process to the continuum, we introduce effective 
wave functions and their integrals, in analogy with the 
$\beta$-decay study of the $^6$He halo nucleus into the $\alpha+d$ continuum \cite{TBD06}. We restrict the presentation to the dominant spin part. 
For the initial $0^+$ state, let us define the effective wave function with hypermomentum $K$
\beq
 u_{\rm eff}^{(K)}(R) =\left(\frac{A-2}{A}\right)^{3/4} R \int dr \, \phi^{00}_{K}(\alpha) \, \frac {{\chi}^{0^+}_{0000K}(\rho)} {\rho^{5/2}} r u_d(r),
\label{Effun}
\eeq
and the effective integrals
\beq
I_E^{(K)}(R)&=&\int_0^R dR' \, u_E(R')u_{\rm eff}^{(K)}(R'),
\nonumber\\
I_E(R)&=&\sum_K I_E^{(K)}(R),
\label{Efint}
\eeq
where $\rho$ and $\alpha$ depend on $(r,R)$, as given in Eq.~(\ref{eq207}).
The normalization factor in Eq.~(\ref{Effun}) arises from the Jacobian between the $(\ve{R},\ve{r})$
and $(\ve{x},\ve{y})$ coordinates.
The reduced matrix elements of the M1 operator (spin part) are then directly proportional to 
$I_E(R)$ (see Appendix B).

\begin{figure}[h]
\begin{center}
\includegraphics[width=10cm,clip=]{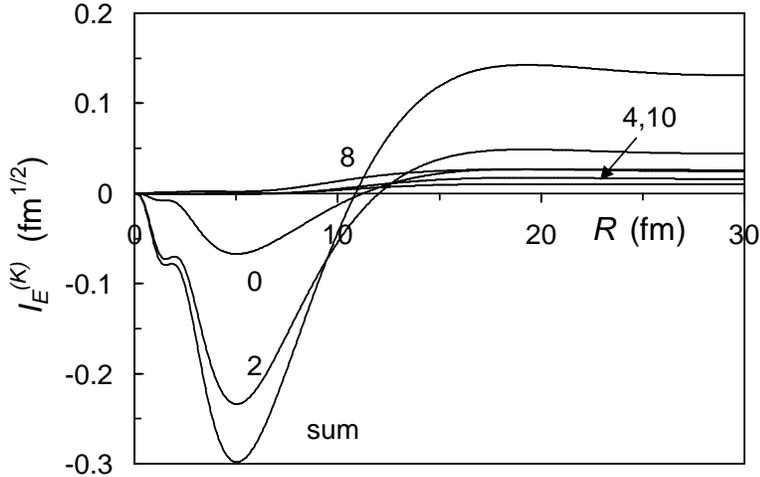}
\end{center}
\caption{Integrals $I_E^{(K)}(R)$  [Eq.(\ref{Efint})] at $E=1$ MeV for the $\alpha+d$ potential $V_a$ and different $K$ values (labels).  
\label{Fig1}}
\end{figure}

In the following, we analyze two properties: the convergence of the hypermomentum expansion, and the
sensitivity of the effective integrals with respect to the $\alpha+d$ potential. Let us start
with the influence of $K_{\rm max}$.
In Fig.~\ref{Fig1} we show the integrals $I_E^{(K)}(R)$ calculated at $E=1$ MeV with potential $V_a$,
for different $K$-values.
The dominant contribution at large $R$ values comes from the $K=0,2,8$ components in the $^6$Li$(0^+)$ wave function. The components $K=4$ and $K=10$ give smaller and
comparable effects to the process. The contributions of other components are small and not 
visible at the scale of the figure. Similar results were obtained for the $\beta$ decay 
of $^6$He \cite{TBD06}. In Ref.~\cite{TBD06} it was shown that the $K=4$ and $K=6$ 
contributions are affected by cancellation effects, which do not occur for 
$K\geq 8$. The situation is therefore very close to 
the $^6$He$(0^+)$ beta decay \cite{TBD06} into the $\alpha+d$ continuum which confirms the halo 
structure of the $^6$Li$(0^+)$ state, suggested by its large r.m.s. radius.

\begin{figure}[thb]
\begin{center}
\includegraphics[width=10cm]{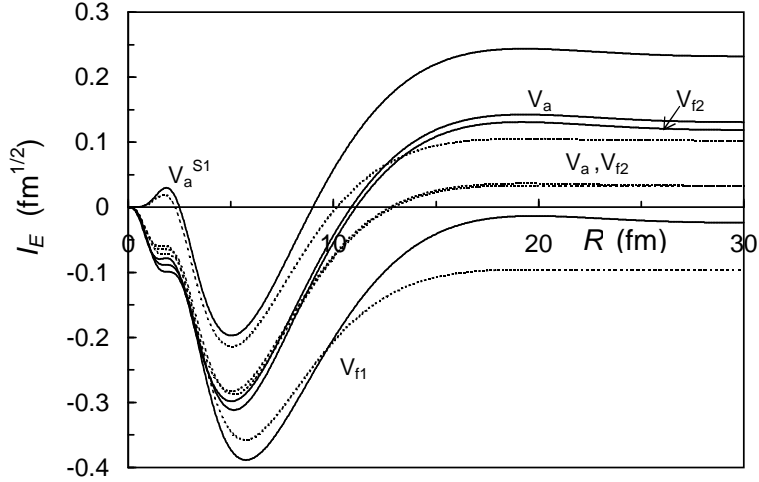}
\end{center}
\caption{Effective integrals $I_E(R)$  [Eq.(\ref{Efint})] at $E=1$ MeV for different $\alpha+d$ potentials (solid lines). Dotted lines represent the equivalent integrals for the $^6$He
$\beta$ decay \protect\cite{TBD06}.
\label{Fig2}}
\end{figure}

In the second step, we analyze the sensitivity of the effective integrals with respect to the 
potential. In Fig.~\ref{Fig2}, these integrals are shown
at $E=1$ MeV. The potentials $V_a$ and $V_{f2}$, which provide similar phase shifts and wave
functions, give results close to each other. 
This is due to the similar node positions near 5 fm of the corresponding scattering wave functions. 
The folding potential $V_{f1}$, owing to a poor phase-shift description, yields a scattering 
wave function with an inner node shifted to the right (about 0.7 fm), and therefore provides 
a different integral. 

In Fig.~\ref{Fig2}, we also show as dotted lines, for each potential, the effective integrals
obtained for the $^6$He $\beta$ decay~\cite{TBD06} (notice that in Ref.~\cite{TBD06}, the factor
$((A-2)/A)^{3/4}=0.74$ in the effective wave function was missing). In that work, 
we have shown that 
a node in the $\alpha+d$ continuum wave functions is responsible for a nearly perfect
cancellation effect in the $\beta$-decay matrix element. 
This is illustrated in Fig.~\ref{Fig2}: for the recommended potential $V_a$, the internal contribution
to the matrix element is about $-0.30$ whereas the external term is about $+0.35$. The final result
is therefore much lower than each component individually. This phenomenon yields a strong sensitivity of
the $\beta$ spectrum with respect to the $\alpha+d$ potential.

Coming back to the $\gamma$ decay of the $^6$Li analog level (solid lines in Fig.~\ref{Fig2}),
the internal parts of the matrix elements are very close to their $^6$He counterparts 
up to about 10 fm. However,
as the long-range parts of the wave functions are different in both nuclei, 
the external contribution to the
$\gamma$-decay matrix element is significantly larger (about $+0.42$ for $V_a$). Consequently,
a cancellation effect still occurs, but is less important. If we disregard potential $V_{f1}$
which does not reproduce the $\alpha+d$ phase shifts, and hence the correct
location of the nodes in the continuum wave functions, all potentials provide
the same sign for the matrix element.

\subsection{M1-transitions to the $\alpha+d$ continuum: transition probabilities}
In Table ~\ref{table2} we give the contributions of different $K$ values to the 
M1 reduced matrix element into the $\alpha+d$ continuum. As we noted above, the orbital and 
spin parts of the M1 transition operator yield nonzero matrix elements only with the
$\ell_x=\ell_y=L=S=1$ ($P$-wave) and $\ell_x=\ell_y=L=S=0$ ($S$-wave) components of the three-body wave function, respectively. As expected from the previous analysis, the dominant contributions come from the $K=0, 2$ 
and 8 components. Additionally, the contribution of the orbital part of the M1 transition operator 
is strongly suppressed (2\% at most).
    
\vspace*{0.5cm}
\begin{table}[h]
\caption{Contribution of different $^6$Li$(0^+$) hypermomenta to the M1  reduced matrix 
elements for transitions into the $\alpha+d$ continuum (in $10^{-3}\mu_N$) for the 
orbital $(L)$ and  spin $(S)$ terms at several energies}
\begin{tabular}{ccccccc} 
\hline
     &\multicolumn{2}{c}{$E=0.5$ MeV}& \multicolumn{2}{c}{$E=1$ MeV}    & \multicolumn{2}{c}{$E=1.5$ MeV}        \\
   $K$ & $(L)$     &   $(S$)   &  $(L)$   & $(S)$    & $(L)$       & $(S)$     \\   \hline
   0 & 0       &-56.9   & 0      &-54.2  & 0         &-46.4   \\ 
   2 & 0.5   &-85.9  & 0.6  &-104.9 & 0.4       &-115.6   \\
   4 & 3.1   &-36.3  & 4.5    &-31.0  & 5.2       &-21.0   \\
   6 & 1.5   &-11.4  & 2.1    &-4.9  &2.2   & 1.4    \\
   8 & -1.2   &-51.1   & -1.6   &-61.6  & -1.7      &-61.3   \\
  10 & -0.3   &-22.2   & -0.3  &-23.1  & -0.3      &-20.0   \\
$>10$& 0.3   &-21.2   & 0.4    &-14.5  & 0.4       &-7.3   \\
Sum & 3.9    &-285.0  & 5.7    &-294.2 & 6.2       &-270.2   \\
\hline
\end{tabular}
\label{table2}
\end{table}
\vspace*{0.5cm}

\begin{figure}[h]
\begin{center}
\includegraphics[width=10cm,clip=]{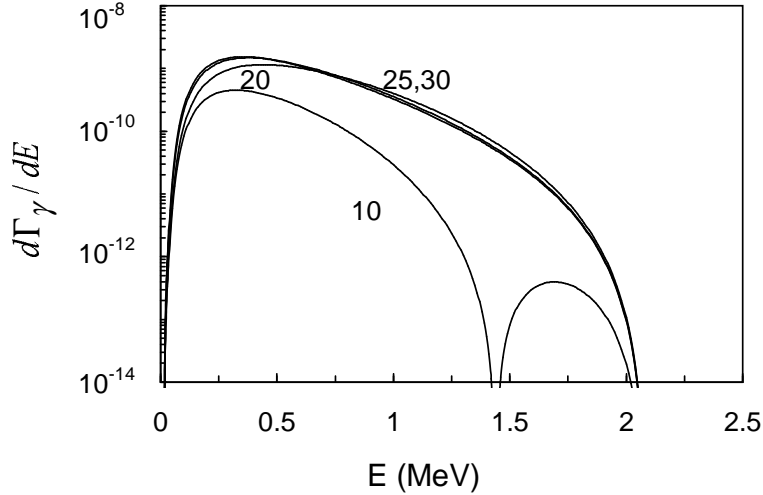}
\end{center}
\caption{Differential width for M1 transitions  into the 
$\alpha+d$ continuum with the $\alpha+d$ potential $V_a$  for several values of $R_{\rm max}$
(in fm). 
\label{Fig3}}
\end{figure}

To analyze the convergence  with respect to the upper bound $R_{\rm max}$ [see Eq. (\ref{Efint})], we display 
in Fig.~\ref{Fig3} the differential 
width $d\Gamma_{\gamma}/dE$  for several values of $R_{\rm max}$ (potential $V_a$ is used).
From Fig.~\ref{Fig3}, one can see that $R_{\rm max}=10$ fm is far from sufficient. 
Achieving a precise convergence requires
larger values ($\sim 25-30$ fm), as in the beta-decay calculations 
of the $^6$He halo nucleus into the $\alpha+d$  continuum \cite{TBD06}. This is
not surprising as the halo structure of $^6$Li($0^+$) is even more pronounced (see Fig.~\ref{Fig0}).
   
\begin{figure}[thb]
\begin{center}
\includegraphics[width=10cm]{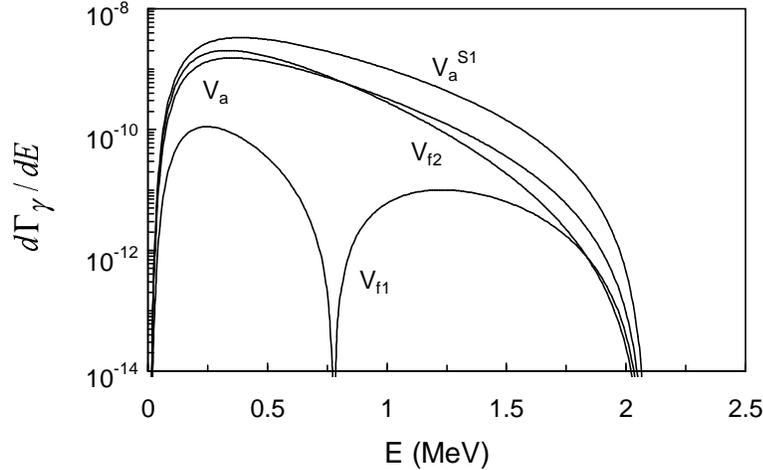}
\end{center}
\caption{Differential width for M1 transitions  into the $\alpha+d$ continuum for several potentials. 
\label{Fig4}}
\end{figure}

In Fig.~\ref{Fig4}, we display the differential width $d\Gamma_{\gamma}/dE$ 
for several $\alpha+d$ potentials.
Contributions from three-body components up to $K_{\rm max}=20$ are taken into account 
with the maximal relative distance $R_{\rm max}=30$ fm. The folding potential
$V_{f1}$ shows a picture strongly different from the other ones, 
with even a sharp minimum at about $E=0.8$ MeV. This potential 
gives a poor description of the $\alpha+d$ phase shift (see Ref.~\cite{TBD06}) and hence a shifted node
position for the $\alpha+d$ scattering wave function. This results in a strong cancellation effect as 
explained in the previous section. The folding potential $V_{f2}$ and the deep potential $V_a$ give 
close results and the supersymmetric potential $V^{S1}_a$ slightly overestimates them.  

The integrated $\gamma$  widths (from $E=0$ up to $Q=2.089$ MeV)
are given in Table \ref{table3} for the different potentials. 
The Gaussian potential $V_a$  simultaneously 
reproduces both the $^6$Li ground state binding energy and the $S$-wave phase shift at low energies. 
Additionally, the $S$-wave scattering wave function of this potential has two nodes at short
distances (one due to the
ground state, and one due to the Pauli forbidden state). 
The nearly phase-equivalent potential $V_{f2}$, which 
also has a forbidden bound state (and hence two nodes at short distances) gives similar results. 

The 
influence of the nodes in the scattering wave function can be tested by using
potential $V^{S1}_a$. The non-physical ground 
state of $V_a$ is removed by using a supersymmetric transformation \cite{Ba87}. 
The resulting phase-equivalent potential $V^{S1}_a$ has exactly the same $^6$Li ground-state energy 
and the same $S$-wave phase shift as $V_a$ but its scattering wave functions have one node less 
at small distances. 
The corresponding width of the M1 transition  is about two times larger (see Table~\ref{table3}).
In the $^6$He $\beta$-decay process, this potential strongly overestimates the data \cite{TBD06}.
Notice that a very different result is obtained with the folding potential $V_{f1}$, 
which has two bound states, but does not reproduce the $\alpha+d$ phase shifts
and the $^6$He delayed $\beta$ decay. 
The shape and magnitude of the transition width and probability are strongly different from the result for $V_a$.  
  
\vspace*{1cm}
\begin{table}[h]
\caption{Integrated  $\gamma$ widths  for different potentials, and branching ratio 
$BR=\Gamma_{\gamma}(0^+ \to \alpha+d)/\Gamma_{\gamma}(0^+ \to 1^+)$ (we use the theoretical value
$\Gamma_{\gamma}(0^+ \to 1^+)=7.5$ eV).}
\begin{tabular}{ccc} 
\hline
  potential         &  $\Gamma_{\gamma}$ (meV) &$BR$  \\ \hline
  $V_a$        & 0.90 & $1.2 \times 10^{-4}$ \\
  $V_{f1}$      & 0.04 & $5.3 \times 10^{-6}$ \\
  $V_{f2}$      & 1.08 & $1.4 \times 10^{-4}$ \\
  $V^{S1}_a$   & 2.27 & $3.0 \times 10^{-4}$ \\
\hline
\end{tabular}
\label{table3}
\end{table}
\vspace*{1cm}

Considering the $V_a$ and $V_{f2}$ potentials, which are consistent with the data on $^6$He $\beta$ decay,
we deduce a recommended branching ratio of $1.3 \times 10^{-4}$ by averaging both values.
A previous estimate \cite{gri98} of the branching ratio 
$\Gamma_{\gamma}(0^+\rightarrow \alpha+d)/\Gamma_{\gamma}(0^+\rightarrow1^+)$ 
provides $0.8\times 10^{-4}$. This value is close to our results obtained with  
potential $V_a$ \cite{Dub94}, and is also similar to the branching ratio observed in the $\beta$ decay of $^6$He \cite{ABB02}. Such a branching ratio should be observable experimentally.

\section{Conclusions}
In the present work, we have studied the M1 transition process from the $^6$Li$(0^+)$ halo state 
into the $\alpha+d$ continuum and into the $^6$Li$(1^+)$ ground state.
Our goal was twofold: $(i)$ to determine the energy distribution of the
$\gamma$ width for the decay into the continuum, and to analyze its sensitivity with
respect to the $\alpha+d$ potential; $(ii)$ to compare this process with the
$^6$He $\beta$-delayed decay. This comparison is a good tool to test charge symmetry in exotic nuclei.
The $^6$Li$(0^+,1^+)$ states are defined in the three-body hyperspherical formalism. 
The experimental  magnetic moment of the ground state and $\gamma$ width 
of the $0^+,T=1$ state are reproduced with a good accuracy.  

We have shown that the spin-dependent term of the M1 transition operator gives the essential 
part of the matrix elements. 
In order to test the influence of the $^6$Li bound-state wave functions, we have also used the
supersymmetric transform \cite{Ba87} instead of the Orthogonalising Pseudopotential method for 
the removal of forbidden states in the three-body wave functions. 
The results are very similar to the present ones, and were therefore not shown.

In the M1 transition probability, the $K=0$ and $K=2$
components of the three-body wave function provide about 50\% of the
matrix elements; consequently, higher hypermomenta play an important role. The same conclusion
holds in the $^6$He $\beta$ decay into the $\alpha+d$ continuum,
where large $K$ values cannot be neglected.

M1 transitions to the continuum provide a good probe of the halo structure in the $^6$Li$(0^+)$
state. The comparison with the $^6$He $\beta$ decay shows that the inner parts of the matrix
elements are very close to each other, as expected from charge symmetry. However, the halo parts are 
different, owing to the different binding energies, and different charges
of the halo nucleons. 
In $^6$Li, the binding energy is lower, and therefore the asymptotic decrease of the wave
function is slower. Consequently the halo contribution is larger in the $\gamma$-decay matrix element,
and even represents the dominant part.
This leads to the conclusion that charge-symmetry breaking is
rather strong in these processes.
The nearly perfect cancellation effect between short-range and halo contributions observed
in $^6$He $\beta$-decay is less important here, and the sensitivity with respect to
the potential is therefore weaker.
Several $\alpha+d$ potentials were tested. The sensitivity is still important (about a factor of 2), but lower than in the $^6$He $\beta$-delayed decay.

The present branching ratio of about $1.3\times 10^{-4}$ is consistent with the value of 
Ref.~\cite{gri98}, where the authors use a simplified model. The present value is based
on potential $V_a$ which reproduces the $^6$Li binding energy, the $\alpha+d$ low-energy
phase shifts, and provides fair results for the $^6$He $\beta$ decay. It is therefore expected 
to have the same quality for the $^6$Li $\gamma$ decay. An experimental
measurement seems to be possible with current facilities, and would provide,
in combination with the data on $^6$He $\beta$ decay, an important step in
a better understanding of the halo structure in isobaric analog states.

\begin{ack}
This text presents research results of the Belgian program P5/07 on 
interuniversity attraction poles initiated by the Belgian-state 
Federal Services for Scientific, Technical and Cultural Affairs (FSTC).
E.M.T. acknowledges the support of the National Fund for Scientific Research (FNRS), Belgium. 
E.M.T thanks the PNTPM group of ULB for its kind hospitality during his stay in Brussels. 
\end{ack}

\newpage
\renewcommand{\appendix}{\par
\setcounter{section}{0}
\setcounter{equation}{0}
\def\thesection{\Alph{section}}
\def\theequation{\thesection.\arabic{equation}}}
\appendix
\section{Gamma delayed transition probabilities to continuum states}
Let us assume a bound initial state at energy $E_i$ with spin and parity $J_i,\pi_i$ of a nucleus
at rest, decaying to a final unbound state at relative energy $E$, and with spin and parity $J_f,\pi_f$. 
In the final state, both nuclei are characterized by spins $I_1$ and $I_2$, and by
internal wave functions $\phi^{I_1}$ and $\phi^{I_2}$.
According to Ref.~\cite{RT67}, the transition probability per time
unit is given by
\begin{eqnarray}
dW_{\gamma}=\frac{2\pi}{\hbar}\frac{|T_{fi}|^2}{(2\pi\hbar)^6} d\ve{p}\, d\ve{P}\, d\ve{p_{\gamma}}\,
\delta (\ve{P}+\ve{p_{\gamma}})\delta (E+E_{\gamma}-E_i),
\label{A.1}
\end{eqnarray}
where we neglect recoil effects. In (\ref{A.1}), ($\ve{p},\ve{P},\ve{p_{\gamma}})$ are the relative, total, and photon momenta. The transition matrix element $T_{fi}$ is obtained from
\begin{eqnarray}
|T_{fi}|^2=\frac{1}{2J_i+1} \frac{2\pi\hbar c}{k_{\gamma}}
 \sum_{\nu_1,\nu_2,M_i,q}
 |\langle\Psi_f^{\nu_1 \nu_2(-)}(\ve{p})|H_{\gamma}^q|\Psi^{J_i M_i \pi_i}\rangle|^2,
\label{A.2}
\end{eqnarray}
where $(\nu_1,\nu_2)$ are the spin orientations in the exit channel, $H_{\gamma}^q$ is the
electro\-magne\-tic-emission hamiltonian with polarization $q$, and $k_{\gamma}$ is the photon
wave number. The final state is described by an ingoing wave
$\Psi_f^{\nu_1 \nu_2(-)}$ with relative momentum $\ve{p}=(p,\Omega_p)$, 
related to the corresponding outgoing wave $\Psi_f^{\nu_1 \nu_2(+)}$ by
\beq
\Psi_f^{\nu_1 \nu_2(-)}(\ve{p})=(-1)^{I_1+I_2-\nu_1-\nu_2}K\Psi_f^{-\nu_1 -\nu_2(+)}(-\ve{p}),
\eeq
where $K$ is the time-reversal operator. The outgoing wave function is written in a partial
wave expansion as
\beq
\Psi_f^{\nu_1 \nu_2(+)}(\ve{p})&=&\sum_{JM\pi\ell I \nu} \langle I_1 I_2 \nu_1 \nu_2|I\nu\rangle
\langle\ell I m \nu|JM\rangle
\Psi^{JM\pi}_{\ell I}(E){\cal D}^{\ell *}_{0 m}(\Omega_p),
\eeq
where ${\cal D}^{\ell}_{0m}(\Omega_p)$ are Wigner functions.
When the relative coordinate $\ve{r}$ is large, the asymptotic behaviour of the partial wave is given by
\begin{eqnarray}
\Psi^{JM \pi}_{\ell I }(E) &\longrightarrow&
\frac{[\pi(2\ell+1)]^{1/2}}{kr}
i^{\ell+1}\exp(i\sigma_{\ell})
 \left( I_{\ell}(kr)-U^{J \pi}O_{\ell}(kr) \right)
\nonumber \\
&&\times [[\phi^{I_1} \otimes \phi^{I_2}]^I \otimes Y_{\ell} (\Omega_r)]^{JM},
\label{A.5}
\end{eqnarray}
where $\sigma_{\ell}$ are the Coulomb phase shifts, and $I_{\ell}$ and $O_{\ell}$ are the ingoing
and outgoing Coulomb functions, respectively.
Here and in the following, we assume a single-channel problem or, in other words, that the dimension
of the collision matrix $\ve{U}$ is unity.

After integration over $\ve{P}$ and $p_{\gamma}$, Eq.~(\ref{A.1}) is transformed as
\begin{eqnarray}
dW_{\gamma}=\frac{k_{\gamma}^2}{(2\pi\hbar)^5 c} |T_{fi}|^2\, d\ve{p}\, d\Omega_{\gamma}.
\label{A.3}
\end{eqnarray}
First, we expand $H_{\gamma}^q$ in electric $(\sigma=E)$ and magnetic $(\sigma=M)$ multipoles
\cite{RB67}. Then
we integrate over the orientations $\Omega_p$  and $\Omega_{\gamma}$. We have
\begin{eqnarray}
\int |T_{fi}|^2 d\Omega_p d\Omega_{\gamma}&=&\frac{32\pi^2}{2J_i+1} 
 \sum_{\sigma \lambda J_f \pi_f }
 \frac{|\alpha_{\lambda}^{\sigma}|^2}{2\lambda+1}\nonumber\\
 && \times \frac{2J_f+1}{2\ell_f+1}
 |\langle\Psi^{J_f \pi_f}_{\ell_f I_f}(E)||{\cal M}_{\lambda}^{\sigma}||\Psi^{J_i \pi_i}\rangle|^2,
\label{A.4}
\end{eqnarray}
where ${\cal M}_{\lambda}^{\sigma}$ are the multipole operators of order $\lambda$ (coefficients $\alpha_{\lambda}^{\sigma}$ are 
given, for instance, in Ref.~\cite{RB67}).
Let use define
\begin{eqnarray}
\Gamma_{\gamma}(E)&=&\sum_{\sigma \lambda J_f \pi_f }
\frac{8\pi k_{\gamma}^{2\lambda+1}}{\lambda (2\lambda+1)!!^2}
\frac{2J_f+1}{2J_i+1}
|\langle\Psi^{J_f \pi_f}_{\ell_f I_f}(E)||{\cal M}_{\lambda}^{\sigma}||\Psi^{J_i \pi_i}\rangle|^2.
\label{A.6}
\end{eqnarray}
Using (\ref{A.4}) in (\ref{A.3}) gives
\begin{eqnarray}
\frac{dW_{\gamma}}{dE}=\frac{\mu k}{2\pi^2 \hbar^3}
\frac{\Gamma_{\gamma}(E)}{2\ell_f +1} ,
\label{A.7}
\end{eqnarray}
where $\mu$ is the reduced mass.
An interesting case concerns transitions to a narrow resonance with energy $E_R$ and particle
width $\Gamma$. In such a case, the scattering wave function can be approximated as \cite{BD85}
\begin{eqnarray}
\Psi^{J_f \pi_f}_{I_f \ell_f}(E) \approx
\frac{1}{k} \frac{[\pi\hbar v (2\ell_f+1) \Gamma]^{1/2}}{E_R-E-i\Gamma/2}
\Psi^{J_f \pi_f}_{BSA},
\label{A.8}
\end{eqnarray}
where $\Psi^{J_f \pi_f}_{BSA}$ is the bound-state approximation of the wave function,
and $v$ the relative velocity. Using this
approximation in (\ref{A.7}) and integrating over $E$ gives
\begin{eqnarray}
W_{\gamma}=\Gamma_{\gamma}^{BSA}/\hbar,
\label{A.9}
\end{eqnarray}
where $\Gamma_{\gamma}^{BSA}$ is the $\gamma$ width in the bound-state approximation. This result
corresponds to the usual definition of the transition probability between two bound states.

\renewcommand{\appendix}{\par
\setcounter{equation}{0}
\def\thesection{\Alph{section}}
\def\theequation{\thesection.\arabic{equation}}}
\appendix
\section{Matrix elements of the M1 transition operator in hyperspherical coordinates}
Let us write the three-body wave function (\ref{eq210}) as
\beq
\Psi^{JM\pi}_{^6{\rm Li}}(\rho,\Omega_5) =  \rho^{-5/2}
\sum_{\gamma K} {\chi}^{J\pi}_{\gamma K}(\rho) 
{\cal Y}^{JM}_{\gamma K}(\Omega_5)=\sum_{\gamma K}\Psi^{JM\pi}_{\gamma K}(\rho,\Omega_5),
\label{eqb1}
\eeq
where index $\gamma$ stands for $(\ell_x \ell_y LS)$. A reduced matrix element of $\ell_x$ is obtained from
\beq
\langle \Psi^{J\pi}_{\gamma K}||\ell_x || \Psi^{J'\pi'}_{\gamma' K'} \rangle& =&
\delta_{\ell_x \ell'_x}\delta_{\ell_y \ell'_y}\delta_{S S'}\delta_{K K'}
[\ell_x(\ell_x+1)]^{1/2}{\hat \ell_x} {\hat L}{\hat L'} {\hat J'} \nonumber \\
&& \times (-)^{\ell_x+\ell_y+S+L+L'+J'} 
\left\{ \begin{array}{ccc}  
L & \ell_x &\ell_y \\
l_x & L' & 1 
\end{array} \right\}
\left\{ \begin{array}{ccc}  
L & J& S \\
J'& L' & 1 
\end{array} \right\} I_{\rho},
\label{eqb2}
\eeq
where we use the notation ${\hat \ell}=\sqrt{2\ell+1}$, and where the integral $I_{\rho}$
is defined as
\beq
I_{\rho}&=&\int {\chi}^{J\pi}_{\gamma K}(\rho)  {\chi}^{J'\pi'}_{\gamma' K'}(\rho) d\rho.
\eeq

Matrix elements of $\ell_y$ are obtained by swapping $\ell_x$ and $\ell_y$. For the crossed term in
(\ref{M1oper}), the calculation is more tedious. We have
\beq
&&\langle \Psi^{J\pi}_{\gamma K}||\ve{x}\times \ve{p_y}+ \ve{y}\times \ve{p_x} || \Psi^{J'\pi'}_{\gamma' K'} \rangle =\delta_{S S'}(-)^{L+S+J'+\ell_x+\ell_y}
\sqrt{6}{\hat \ell_x}{\hat \ell_y}{\hat \ell_x'}{\hat \ell_y'}{\hat L}{\hat L'} {\hat J'}\nonumber \\
&& \times 
\left( \begin{array}{ccc}  
\ell_x'& 1 &\ell_x \\
0& 0 & 0 
\end{array} \right)
\left( \begin{array}{ccc}  
\ell_y'& 1 &\ell_y \\
0& 0 & 0 
\end{array} \right)
\left\{ \begin{array}{ccc}  
L & J& S \\
J'& L' & 1 
\end{array} \right\}
\left\{\begin{array}{ccc}
         \ell_x & \ell_y & L \\
         \ell_x'& \ell_y' &L'\\
        1 &    1   &   1     \\ 
\end{array}\right\} I_{\rho}\, I_{\alpha},
\label{eqb4}
\eeq
where the angular integral reads
\beq
I_{\alpha}&=&\int_0^{\pi/2} d\alpha \cos^2\alpha \sin^2\alpha
{\phi}^{\ell_x \ell_y}_{K}(\alpha)  
\left( \frac{d}{d\alpha}+\frac{\Delta l_y}{\tan \alpha}-\frac{\Delta l_x}{\cot \alpha} \right)
{\phi}^{\ell'_x \ell'_y}_{K'}(\alpha).
\label{eqb5}
\eeq

In this expression, $\Delta \ell=1+[\ell'(\ell'+1)-\ell(\ell+1)]/2$. Integration over $\alpha$ is performed numerically.
For the hyperradius $\rho$, the use of Lagrange functions makes the integral very simple.

For the spin part of the M1 operator, we have
\beq
\langle \Psi^{J\pi}_{\gamma K}||\ve{s}_1 || \Psi^{J'\pi'}_{\gamma' K'} \rangle &=&
\delta_{\ell_x \ell'_x}\delta_{\ell_y \ell'_y}\delta_{L L'}\delta_{K K'}
{\hat s_1} {\hat S}{\hat S'}{\hat J'}  [s_1(s_1+1)]^{1/2} \nonumber \\
&& \times (-)^{s_1+s_2+L-J}
\left\{ \begin{array}{ccc}  
J & S& L \\
S'& J' & 1 
\end{array} \right\}
\left\{ \begin{array}{ccc}  
S & s_1& s_2 \\
s_1& S' & 1 
\end{array} \right\}
I_{\rho},
\label{eqb6}
\eeq
where we have assumed that the core spin is zero ($s_3=0$).

For transitions to the continuum, the previous formula can still be applied, but the final-state
wave functions are now defined by Eq.~(\ref{eq200}).
It is clear that with the restriction to the $S$-wave final state, the orbital components 
$\ell_{x,\mu}$ and $\ell_{y,\mu}$ do not contribute to the M1 transition. 
The matrix element of the crossed term is performed over the Jacobi coordinates. Using the $S$-wave
character of the scattering state, we have
\beq
&&\langle \Psi^{0^+}_{^6{\rm Li}}||\ve{x}\times \ve{p_y}+ \ve{y}\times \ve{p_x} || \Psi^{1^+}_{\alpha d} \rangle
=\sqrt{2\mu_{pn}\mu_{\alpha d}/9}\sum_K \int dx dy \phi^{11}_{K}(\alpha) {\chi}^{0^+}_{1111K}(\rho)
\nonumber \\
&&\times xy\rho^{-5/2}\left(x\frac{\partial}{\partial y}-  y\frac{\partial}{\partial x} \right)
u_d(x/\sqrt{\mu_{pn}})u_E(y/\sqrt{\mu_{\alpha d}}) ,
\label{eqA21}
\eeq
where $\rho$ and $\alpha$ are given in Eq.~(\ref{eq207}).
The spin contribution is obtained with the same technique, with the help of Eq.~(\ref{eqb6}). Note
that the bra and ket have been swapped with respect to Eq.~(\ref{eq204}). The ordering is simply
restored with a factor $-1/\sqrt{3}$.

\end{document}